\documentclass[PHME, 2024]{PHMSociety}
\usepackage{graphicx}
\usepackage{amsmath}
\usepackage{siunitx}
\begin{document}

% Paper Title
\title{Testing Topological Data Analysis for Condition Monitoring of Wind Turbines}

% Authors List
\author{%			
	Simone Casolo\authorNumber{1}, Alexander Johannes Stasik\authorNumber{2}, Zhenyou Zhang\authorNumber{3}, and Signe Riemer-Sørensen\authorNumber{4}
}

% Author Affiliations
\address{% This is a tabular environment so each affiliation needs to be separated by "\\" or "\tabularnewline"
	\affiliation{{1}}{Cognite AS, Oslo, Norway}{ %add emails
		{\email{simone.casolo@cognite.com}}
		} % emails input
	\tabularnewline % skip one row for next affiliation	
	\affiliation{2, 4}{Sintef Digital, Oslo, Norway}{ % add emails
		{\email{alexander.stasik@sintef.no}}\\
        {\email{signe.riemer-sorensen@sintef.no}}
		} % emails input
  	\tabularnewline % skip one row for next affiliation	
	\affiliation{3}{ANEO AS, Trondheim, Norway}{ % add emails
		{\email{zhenyou.zhang@aneo.com}}
		} % emails input
}

% Create the title
\maketitle
\pagestyle{fancy}
\thispagestyle{plain}

% PHM Society Distribution License Information, provide first author's name "FirstName LastName"
% NOTE: Do not forget to add this. The paper will not be accepted without this open access license footnote.
\phmLicenseFootnote{Simone Casolo}

% Abstract
\begin{abstract}%   %NOTE: Deleting the percentage after "{abstract}" may be lead to an extra leading space in the first line of the abstract, and this should be prevented.
% Abstracts are required for all papers, and an abstract of 150-400 words should be included at the beginning of the paper. The abstract should be formatted as an unnumbered section and it is preferred to be presented in a single paragraph. Define all symbols and expand all abbreviations used in the abstract. Do not cite references in the abstract.

We present an investigation of how topological data analysis (TDA) can be applied
to condition-based monitoring (CBM) of wind turbines for energy generation.\\
TDA is a branch of data analysis focusing on extracting meaningful information from complex datasets by analyzing their structure in state space and computing their underlying topological features. By representing data in a high-dimensional state space, TDA enables the identification of patterns, anomalies, and trends in the data that may not be apparent through traditional signal processing methods.\\
For this study, wind turbine data was acquired from a wind park in Norway via standard vibration sensors at different locations of the turbine's gearbox. Both the vibration acceleration data and its frequency spectra were recorded at infrequent intervals for a few seconds at high frequency and failure events were labelled as either gear-tooth or ball-bearing failures. The data processing and analysis are based on a pipeline where the time series data is first split into intervals and then transformed into multi-dimensional point clouds via a time-delay embedding. The shape of the point cloud is analyzed with topological methods such as persistent homology to generate topology-based key health indicators based on Betti numbers, information entropy and signal persistence. Such indicators are tested for CBM and diagnosis (fault detection) to identify faults in wind turbines and classify them accordingly. Topological indicators are shown to be an interesting alternative for failure identification and diagnosis of operational failures in wind turbines.

\end{abstract}

\section{Introduction}
\label{sec:intro}
\begin{figure}[t]
\centering
\includegraphics[width=.99\columnwidth]{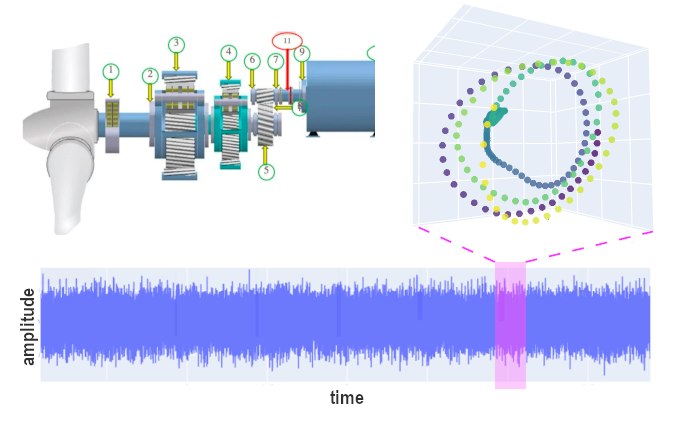}
\caption{Overview of how the gearbox vibration
data are processed by means of topological data analysis.}
\label{fig:overview_figure}
\end{figure}

The global demand for renewable energy sources has seen a significant rise in recent decades, with wind energy emerging as a prominent contributor to sustainable power generation \cite{RevWind1}. 
Wind turbines, pivotal in harnessing wind energy, operate under
diverse environmental conditions and mechanical stresses, making their maintenance and monitoring crucial for optimal performance and longevity. Condition-based monitoring (CBM) has emerged as a proactive approach to monitor the health of wind turbines, aiming to detect faults and predict potential failures before they escalate, thus minimizing downtime and maintenance costs \cite{RevWind2}.\\
Traditional CBM methods often rely on spectral signal processing techniques to analyze sensor data for anomaly detection and fault diagnosis.  Signal analysis techniques are commonly used for fault diagnosis and typically apply tools such as Fourier or wavelet analysis of frequency signatures from accumulated time series generated from sensors installed on wind turbines. Where possible, machine learning techniques are then used to identify early signatures of failure in the data and alert engineers as soon as the equipment's health starts deteriorating. However, frequency-based methods often require accumulating signals 
for a significant time before processing them successfully, making it an ideal method for analyzing failures after they occur. Online fault detection is much more challenging, and together with inherent complexity and non-linearity in wind turbine data, pose challenges for conventional analytical approaches.\\
To address these challenges, alternative data analysis techniques
have gained attention for their ability to extract meaningful insights
from complex datasets. Among those, topological data analysis (TDA) has
recently risen as a possible alternative. 
TDA is a branch of data analysis that focuses on revealing the 
the underlying structure of datasets by analyzing their shape:  
particularly their topology in high-dimensional state spaces. 
By representing data as multidimensional point clouds 
and leveraging mathematical tools from algebraic topology, TDA enables the identification of intricate patterns, anomalies, and trends that may not be discernible through traditional signal processing methods alone (see Fig.\ref{fig:overview_figure}).\\
In this study, we explore how TDA techniques can be employed to analyze vibration data collected from wind turbines at a wind park. Vibration sensors placed strategically in different locations of the turbine's gearbox provide high-frequency data capturing both vibration acceleration and frequency spectra.By employing a systematic data pipeline, including time-series segmentation and time-delay embedding, we transform the raw sensor data into a multidimensional point cloud and then, process it via topological analysis.\\
The primary objective of this research is to evaluate how topological indicators derived from TDA, such as Betti numbers, information entropy, and signal persistence can be used or complement more traditional spectral analysis as key health indicators for CBM and fault diagnosis in wind turbines. 

\section{Data description}
For this analysis, we use vibration data collected from two wind turbine gearboxes from a wind park
located in Norway.
The data sets are proprietary, owned by the wind park operator ANEO (www.aneo.no)
and this work is the first publicly available analysis of the data.
The data was collected using accelerometers, located at various positions in the gearbox.
For the analysis, we focused on sensors that were physically closest to the known failure
positions and most correlated with the time of failure of the gearbox. The considered
sensors are located at the gearbox high-speed stage front (GbxHssFr), at the gearbox
intermediate stage (GbxIss), at the gearbox planetary stage (Gbx1Ps), and at the non-drive
end of the generator (GnNDe). The left panel in Figure \ref{fig:example-data} show a 
\SI{0.05}{s} example of vibrations recorded from GbxHssFr.
The vibration / acceleration data were sampled at 25.6 kHz for 10 seconds at infrequent intervals.
The two cases have respectively 23 and 21 samples of 10 s length with a sampling
rate of 25.6 kHz. The data is collected at infrequent intervals over approximately
a year until the time when
failures happened, and the equipment was stopped for maintenance. 
In the first case, data were acquired from  
2022-10-28 to 2023-10-11 and data ended with a ball bearing failure (BBF) at
the non-drive end of the generator.
In the second case, data was recorded from 2022-05-24 to 
2023-06-21 ended with a gear tooth failure (GTF) at the planetary stage section 
of the gearbox.

\section{Methods}
\label{sec:methods}
In this section, we delineate the methodologies employed for analyzing complex data
structures, focusing particularly on spectral analysis and topological data analysis
(TDA). Spectral analysis, rooted in the principles of linear algebra and signal
processing, extracts valuable insights from data by decomposing it into its constituent
frequencies. Conversely, topological data analysis, drawing from the field
of algebraic topology, examines the shape and connectivity of data through the lens of
persistent homology, providing a holistic understanding of its underlying structure.\\
Both spectral analysis and TDA offer distinct yet complementary approaches to
understanding complex datasets. While spectral analysis emphasizes frequency-based
decomposition, TDA highlights the intrinsic topological features of the data. By comparing 
and contrasting these methodologies, we aim to elucidate their respective strengths, 
limitations, and applicability in various analytical contexts. This comparative analysis 
serves as a foundation for our subsequent exploration and interpretation of results, 
contributing to a comprehensive understanding of the dataset under investigation.

\subsection{Spectral analysis}
Spectral analysis, a fundamental technique in signal processing and data analysis, 
provides a powerful framework for decomposing complex data. Rooted in the principles of Fourier series, spectral analysis offers invaluable insights into the underlying structure and dynamics of various data types across diverse domains, including engineering, physics, biology, and finance.\\
At its core, spectral analysis aims to characterize the frequency content of a signal or dataset. By representing data in the frequency domain, analysts can identify dominant patterns, periodicities, and trends that may not be readily apparent in the time or spatial domain. This decomposition facilitates the extraction of meaningful information, enabling researchers to discern underlying patterns, detect anomalies, and make informed predictions. \\
Spectral analysis is a common tool for condition monitoring 
in wind turbines \cite{wind2,wind3}. Vibration data are typically collected 
from sensors placed in correspondence to moving elements in turbine generators and 
gearboxes, subject to wear and mechanical failure. Data are analyzed to 
identify anomalies and expose drift and changes in the data that can be associated
with a degradation of the system health and, inturn, lead to its mechanical failure
\cite{wind1, RevWind1, RevWind2}.\\
One of the key advantages of spectral analysis lies in its ability to unveil hidden relationships and structures within data. Through techniques such as Fourier transform, wavelet analysis, and singular value decomposition (SVD), analysts can disentangle complex signals into simpler components, each representing a distinct frequency or mode of variation. This spectral decomposition forms the basis for a wide range of applications, including signal filtering, noise reduction, feature extraction, and system identification.

\subsection{Topological data analysis}
\begin{figure*}[t]
    \includegraphics[width=\textwidth]{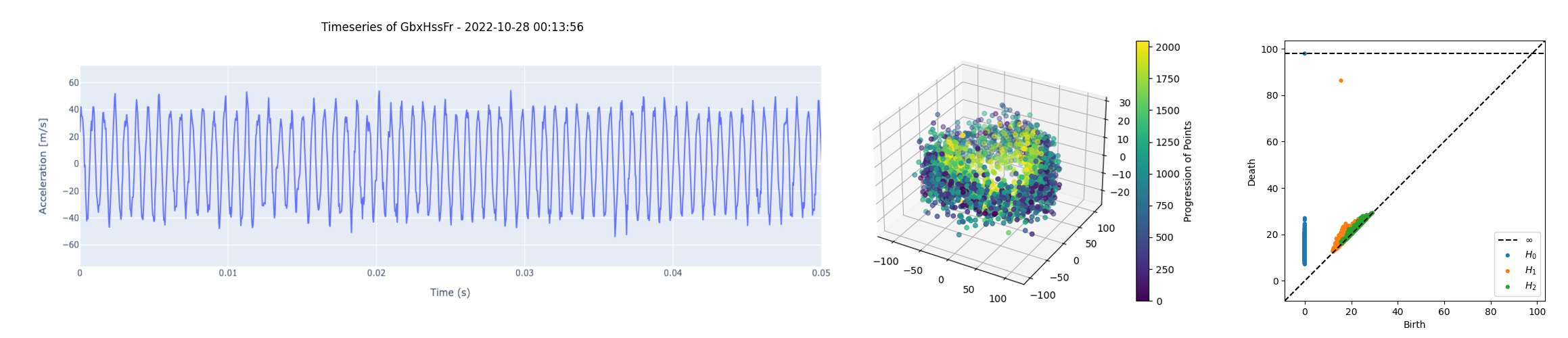}
    \caption{Left to right: Raw time-series signal, embedded point cloud and persistence diagram for GbxHssFr sensor at normal operation state. Note the toroidal point cloud,
    resulting from the embedding of the periodic time series. The loop structure 
    is revealed in the persistence diagram as a point (yellow) far from the diagonal, 
    where points created by signal noise tend to accumulate.}
    \label{fig:example-data}
\end{figure*}

Topological data analysis allows the interpretation of the spatial arrangement 
of data. This approach has been developed in the last decade and successfully applied to the analysis of data in several fields of engineering, fluid mechanics \cite{CasoloTDA}, physics and biology \cite{TDA_Review}. Here we will present a brief introduction to the topic: for a full exposition of this approach, we recommend the excellent articles from  Perea and Harer \cite{PereaHarer}, Chazal et al. \cite{chazal} and Smith et al. \cite{SMITH2021107202}.\\
A common assumption in data analysis is the hypothesis that there exists a suitable space of parameters where data happen to form a manifold. In this case, it would be fair to assume that the shape of such a manifold would contain information about the data. TDA is one of the tools that can be used to interpret such information. Univariate time series of a scalar signal is not immediately suitable to be analysed with TDA. The signal is therefore embedded with a time-delay approach to form a high-dimensional space via a procedure known as Takens embedding\cite{TakensTheo}. This method embeds a time signal into a vector without loss of information, by defining two parameters: the time-delay $\tau$ and the embedding dimension $d$. Then, the time series $\textbf{x}(t)$ is sampled in $d$-points, each separated by a time $\tau$. The embedded $d$-dimensional vector is then built as:
\begin{equation}
\textbf{x}(t) = \{x(t), x(t-\tau), \dots, x(t-d\tau)\}    
\end{equation}
As the time series evolves in time, it can be sampled repeatedly to build
a series of vectors, which are accumulated to form a point cloud in
$d$-dimensions. This cloud samples the manifold on which the data lays.\\
Once the data are represented in the $d$-dimensional space of the embedding,
this can be analyzed by using algorithms developed in algebraic topology. 
To build the manifold, it would be required to connect each vector,
\emph{i.e.} point in the cloud within a given radius around each point, 
to form a network or a cell complex. This process is performed by connecting points lying within a given radius via 
the creation of Vietoris-Rips complexes: a simplicial (cell) complex representing the connectivity between data points in a dataset. 
To encode the complexity of the point 
cloud, we then compute a nested series of complexes that are formed at every point
increasing the value of the radius in a process known as filtration. 
The construction of the complex involves considering all possible subsets 
of data points and connecting those that are within a specified distance
threshold. Overall, the point cloud
generated from the time series is a sampling of the shape of the 
data, and the filtration process generates several simplicial complexes
which are the computational descriptions of the shape of the data.
As the filtration parameter increases, the Vietoris-Rips complex captures 
increasingly complex topological features, ranging from individual points
to higher--dimensional structures such as loops and voids. Typically, these 
features are unique to the data manifold 
\cite{ATTALI2013448} and are the topological structures 
we consider when analyzing the data.\\
The presence of loops, voids, etc. is encoded in the concept of homology. 
Persistent homology analyze the development of data sets by considering the
evolution of topological features across different scales. It quantifies the
persistence of these features as they emerge, merge, or disappear, providing a 
robust framework for capturing and characterizing the essential topological 
structure of complex datasets. Each structure then has 
a birth and a death value at a given radius of the filtration process, 
which can be recorded in a diagram known as a persistence diagram, unique for the 
analyzed shape. Each point in the diagram corresponds to a 
topological feature per each dimension (connected components in dimension 0, loops in 
dimension 1, voids in dimension 2, etc.)
with its coordinates indicating the scale at which the feature is born and 
dies (see Figure \ref{fig:example-data} for an example of a persistence diagram). The persistence of a feature is measured as the difference between 
its death ($d$) and birth ($b$) scales. Naturally, persistence diagrams 
are non-empty only above the diagonal as the death of a feature would occur
only after its birth, and the more 'persistent' a feature is, the further this 
would lay from the diagonal line.\\
By analyzing persistence diagrams, it is possible to 
identify persistent features that are robust across multiple scales and 
distinguish them from transient noise or artefacts in the data.
Topological indicators in each homology dimension $H_k$ can be extracted from 
persistence diagrams and used to analyze data.\\
While TDA can be applied to uncover the shape of the data manifold for a 
signal of an arbitrarily long time, it can also be applied to a sequence of
short time windows, sliding forward in time and partially overlapping 
\cite{PereaHarer}.
This sliding windows approach can be used to 
uncover the local structure of data and their evolution and it has been used
successfully to study the dynamics of mechanical systems.

\subsection{Topology of vibration signals}
Topological methods are expected to work particularly well for analyzing 
periodic time signals and their changes. Mathematically, it can be shown that 
periodic signals which can be approximated with a trigonometric function 
of a given frequency, can be embedded into a point cloud of elliptical 
shape, hence in a loop that should be detected by a high persistence signal 
of dimension 1 ($H_1$) \cite{PereaHarer}. 
When the signal is instead composed of combinations of more frequencies 
these give rise to more complex manifolds such as tori and higher dimensional 
structures \cite{perea2016persistent}.\\
In the case of oscillating systems, according to the Arnol'd-Liouville theorem in 
dynamical system theory, systems of $n$ harmonic oscillators give rise to trajectories 
on a $n$-dimensional torus.  
This phenomenon emerges due to the conservation of action variables, which 
characterize the system's motion in phase space. In a system of harmonic oscillators, 
each oscillator contributes a set of action-angle variables, representing the 
oscillation's amplitude and phase in each dimension. These variables remain constant 
over time, preserving the system's dynamics. As a consequence, trajectories in phase 
space form closed loops, tracing out toroidal surfaces \cite{Arnold}. 
This behaviour stems from the 
periodicity of harmonic motion, enabling the system's state to return to its initial 
configuration after completing a cycle. The toroidal topology of these trajectories 
reflects the periodicity and conservation of action variables, illustrating a 
fundamental principle of dynamical systems theory.\\
When a vibrating mechanical system such as the gearbox of a wind turbine oscillates, 
it is reasonable to expect, accounting for deviation and noise, a behaviour
similar to that of a harmonic oscillator, hence a trajectory in phase space 
spanning a manifold similar to a torus. In this case, it would be reasonable to 
expect some homology signatures that should be visible from the persistence
diagrams, making persistent homology a good candidate method for characterizing the 
dynamics of vibrations at the gearbox and, hopefully, spotting the 
appearance and evolution of abnormal behaviour from sensors' time series.

\subsection{Analysis strategy}
In this work, we have chunks of high-frequency data sparsely collected, each a few weeks or months apart. Every chunk of data is sampled with \SI{25.6}{kHz} for a period of \SI{10}{s}, allowing for spectral, spectral-temporal or topological data analysis. We assume that any changes happen on time scales of days or weeks, and hence the data is stationary over each of those \SI{10}{s} segments. Therefore, the main strategy of our analysis focuses on finding trends
between time segments as we get closer to the failure time.\\
The key challenge in this work is the lack of ground truth, as we do not know the onset of the damage that eventually led to the failure of the gearbox. Therefore, we use the early stages of data as a baseline, assuming that the damage developed later. In other words, we are looking for systematic deviations from the early state which is assumed to be \textit{healthy}.
Topological data analysis was performed with the Giotto-TDA code suite 
\cite{giottoph}. Time series from vibration sensors were embedded using Takens embedding with the optimal time delay and embedding dimension 
chosen by the built-in standard heuristics based on mutual information \cite{MutualInfo, FalseNN}. Persistence diagrams $D$ were then compiled from the Vietoris-Rips complexes obtained from the filtration and used to compute the following topological indicators:\\
{\bf The maximum persistence}, defined as the infinity norm for each homology dimension:
\begin{equation}
\mathcal{P}^{\textrm{H}_k}_{\infty}(D_{\textrm{H}_k}) = 
\max_{\{b,d \}\in D } | d-b |  
\end{equation}
This is a useful shape indicator as noise gives rise 
to points in $D$ with a short lifetime, while relevant 
features of the points cloud 
(\emph{e.g.} loops) are expected to have high persistence.\\
{\bf The normalized persistence entropy} is another measure of complexity \cite{PerEntropy,ATIENZA2020107509},
$\overline{\textrm{E}}_{\textrm{H}_k}(D)$, expressed as a measure
of the distribution of points along the diagram based on Shannon's
entropy formula:
\begin{equation*}
    \overline{\textrm{E}}_{\textrm{H}_k}(D) = 
    - \frac{1}{\log_2 {\mathcal{S}(D)}}
    \sum_{\{b,d \}\in 
    D_{\textrm{H}_k} } \frac{|d-b|}{\mathcal{S}(D)}
    \log_2 \left ( \frac{|d-b|}{\mathcal{S}(D)} \right ) 
\end{equation*}
where the amplitude $\mathcal{S}(D_{\textrm{H}_k})$ for a 
given dimension is defined as:
\begin{equation}
\mathcal{S}(D_{\textrm{H}_k}) = \sum_{\{b,d \}\in D }|d-b|
\end{equation}
{\bf Betti curves} are another informative topological indicator, which measures the amount of $k$-dimensional topological features \emph{i.e.} 
the Betti number, $\beta_k$ \cite{Hatcher2002},
at each value of the filtration parameter.
In practice, these "count" the number of $k$-dimensional 
holes of a space: $\beta_0$
represents connected components, $\beta_1$ circles, $\beta_2$ voids, etc.
As an example, for a two-dimensional circle the set of Betti numbers 
$\{\beta_0,\beta_1,\beta_2\}$ are 
$\{1,1,0\}$, for a filled disk $\{1,0,0\}$, a hollow sphere $\{1,0,1\}$, 
for a filled ball $\{1,0,0\}$, for a torus $\{1,2,1\}$, etc.\\
Other indicators are the {\bf $f$-family of indicators} defined here, as
proposed by Adcock et al. \cite{Carlsson_f} and used in TDA for the 
anomaly detection in rotating equipment for manufacturing
\cite{chatter2, chatter3}
as they combine the highest persistence with amplitude information:
\begin{equation}
\begin{array}{c}
     f_1 = \sum_i b_i \cdot (d_i - b_i) \\
     \\
     f_2 = \sum_i (d_{max} - d_i) - (d_i - b_i) \\ 
     \\
     f_3 = \sum_i b_i^2 \cdot (d_i - b_i)^4  \\
     \\
     f_4 = \sum_i (d_{max} - d_i)^2 - (d_i - b_i)^4
\end{array}
\end{equation}

\section{Data Analysis}
No data cleaning or pre-processing has been performed to the signal prior to the analysis described in Section \ref{sec:methods}, hereafter addressed as 'raw data'. 

\subsection{Bearing Failure}
\begin{figure}[t]
    \includegraphics[width=\columnwidth]{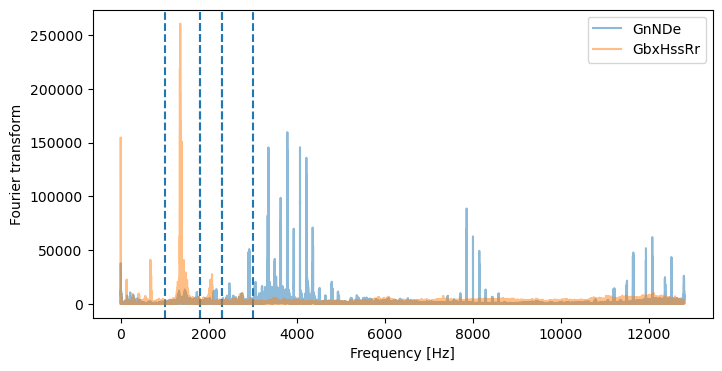}
    \caption{Fourier transform (normalised to counts) of the signal recorded on 2023-10-28 for GnNDe-BBF and GbxHssFr-BBF. The vertical lines indicate the frequency intervals for which the most dominating peaks are investigated for GbxHssFr.}
    \label{fig:FFT}
\end{figure}
\begin{figure}[h]
        \includegraphics[width=\columnwidth]{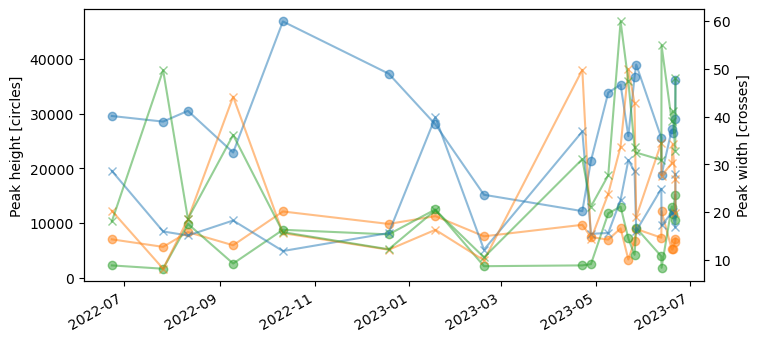}
    \caption{Peak height (left axis, circles) and width (right axis, crosses) for three frequency signatures (most dominant peak in the frequency ranges [1000, 1800], [1800, 2300], [2300, 3000] Hz) for GbxHssFr in the bearing failure case.}
    \label{fig:figu}
\end{figure}
\begin{figure}[t]
    \includegraphics[width=\columnwidth]{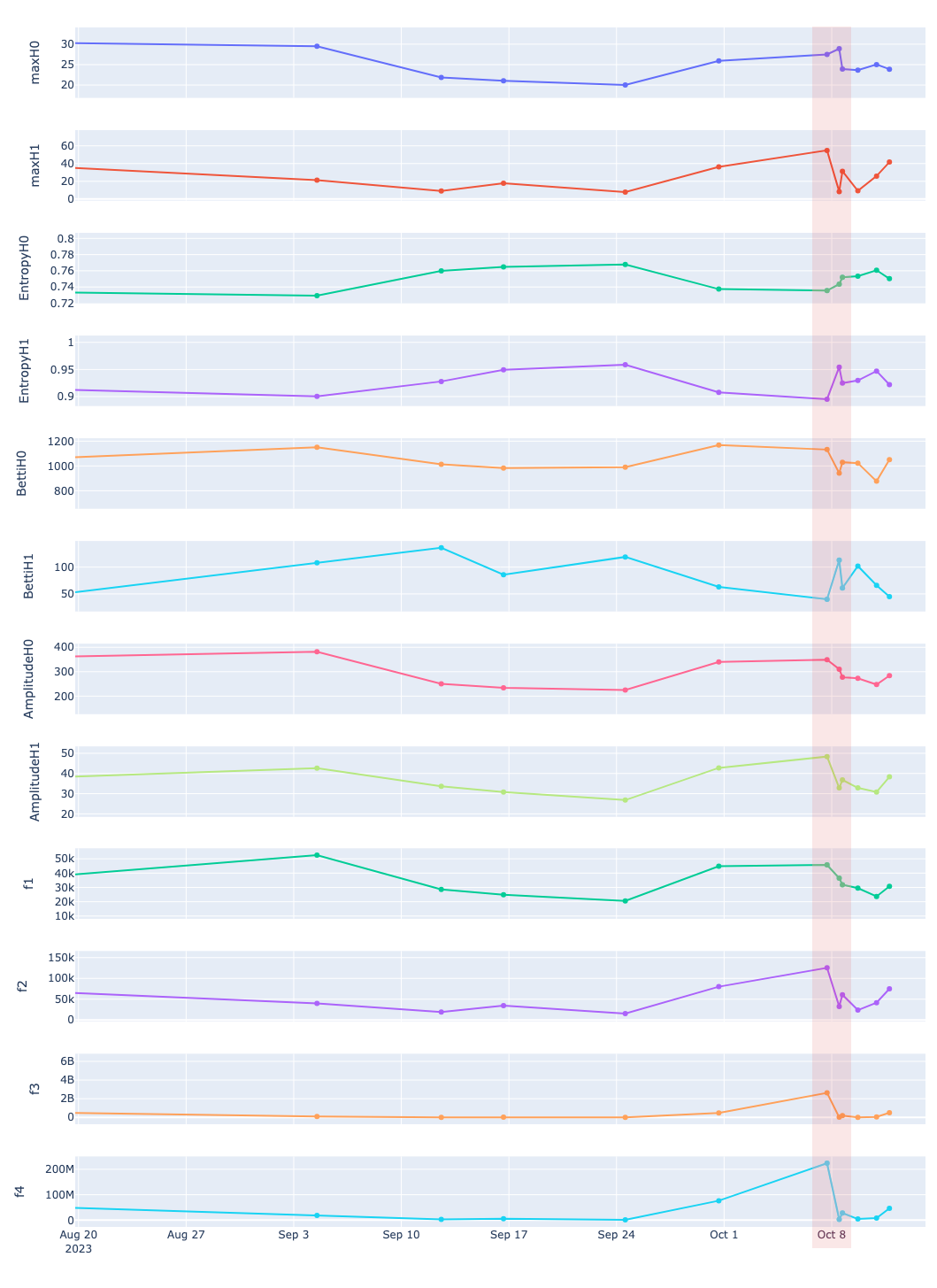}
    \caption{Topological indicators computed for the signal GbxHssFr in the 
    bearing failure case. Highlighted the most significant anomaly, dated 
    2023-10-08.}
    \label{fig:GbxHssFrTDA}
\end{figure}
The bearing failure was reported at the non-drive end of the generator, corresponding to the location of the sensor
labelled as "GnNDe" and the signal was recorded sporadically between October 2022 and the failure on November 11 2023.
Each time series records acceleration data for the sensor and the corresponding frequency spectrum is computed from the raw signal through a Fast Fourier Transform (FFT) approximation. Figure \ref{fig:FFT} shows the spectrum for the signal recorded at the GnNDe (blue) and at the earliest available timestamp, 28-10-2023. We assume this to correspond to a state of "normal operations".\\
Topological analysis shows the point cloud 
corresponding with GnNDe is not describing a torus, but rather a 
semi-uniform ball, indicating non-periodic or very noisy behaviour. As a consequence, the $H_0$ persistence can only be interpreted 
as a measure of how much clustered  or diffused the data are in the parameters space, 
while $H_1$ and higher-dimensional homology signals are expected to be 
low and not significant. Indeed, the only noticeable trend in the
topological indicators is a decrease in $H_0$ persistence and an 
increase in entropy, typically as a consequence of a progressively  
less structured and more noisy signal. At a closer look, other sensor
signals seem more suitable for analysis. In particular, the intermediate and 
high-speed stage sensors (GbxIss and GbxHss, respectively) show a more
periodic and regular behaviour. Indeed the high-speed front (GbxHssFr)
sensor shows a clear oscillating signal and a frequency
spectrum dominated by a peak at around \SI{1400}{Hz} and its multiples (orange spectrum in Figure \ref{fig:figu}). The embedded signal clearly shows a toroidal shape, 
a "filled" torus consisting of one main loop induced by the main frequency
component, and the direction orthogonal to the loop blown up by the noise.
The corresponding persistence diagram then shows a high persistence 
point for $H_0$ and one at $H_1$ corresponding to the loop and proportional 
to its size.\\
The analysis of the evolution of the GbxHssFr signal is not trivial. Figure \ref{fig:FFT} shows the time development of the most dominant peak in each of the three frequency bands shown in 
Figure \ref{fig:figu}. We found that the frequencies do not shift 
significantly until the time of the failure (not shown). 
The corresponding peak heights and widths show a larger spread, 
especially at the lowest frequency.
We also measure the evolution by computing the mutual distance between the vectors containing Fourier
coefficients for each time series. This distance becomes more evident between the signal in the early timestamps (i.e. normal 
operations) and signals in a few specific days close to the failure, 
in particular at 2023-10-08 and 2023-10-10, one and three days from the 
failure, especially for the components included from 0 to \SI{1800}{Hz}.\\
We observe a similar behaviour in the skewness and kurtosis of the raw signal,
which show a slow decreasing trend, with a very high spike in the latter 
at the timestamp 08-10-2023, 3 days from the point of failure, which was 
not evident from the spectra alone. The monitoring
of kurtosis in the early detection 
of bearing failures is well-known in the literature and it is likely 
to be a good indicator in this case as well \cite{kurt3, kurt2, kurt1}.\\
\begin{figure}[t]
    \includegraphics[width=\columnwidth]{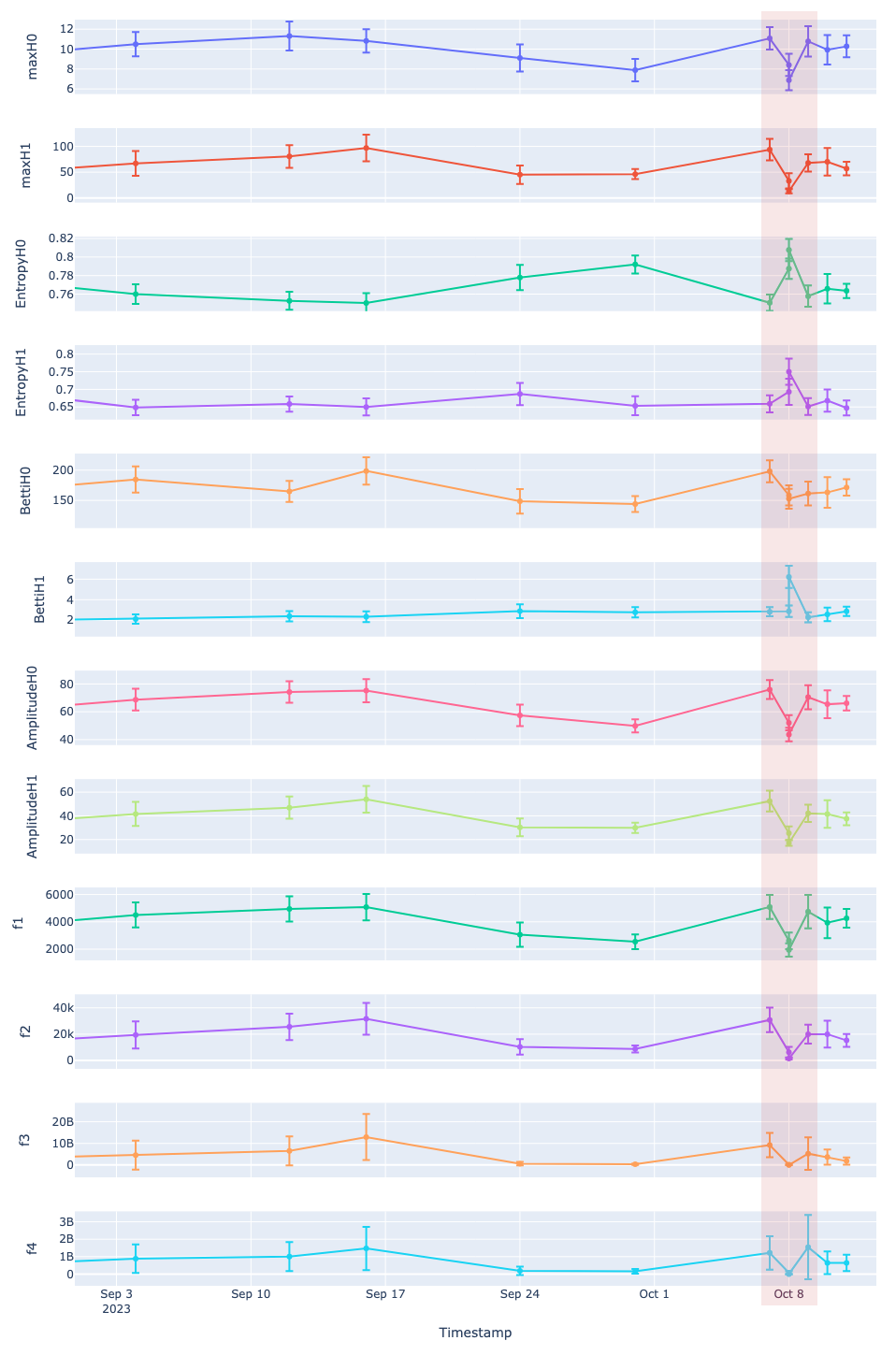}
    \caption{Topological indicators obtained by averaging the results of
    several sliding windows of \SI{5}{ms},  
    computed for each of the chunks for GbxHssFr in the 
    bearing failure case. The most significant anomaly is dated 
    2023-10-08.}
    \label{fig:GbxHssFr_AM_case2}
\end{figure}
\begin{figure*}[t]
    \includegraphics[width=\textwidth]{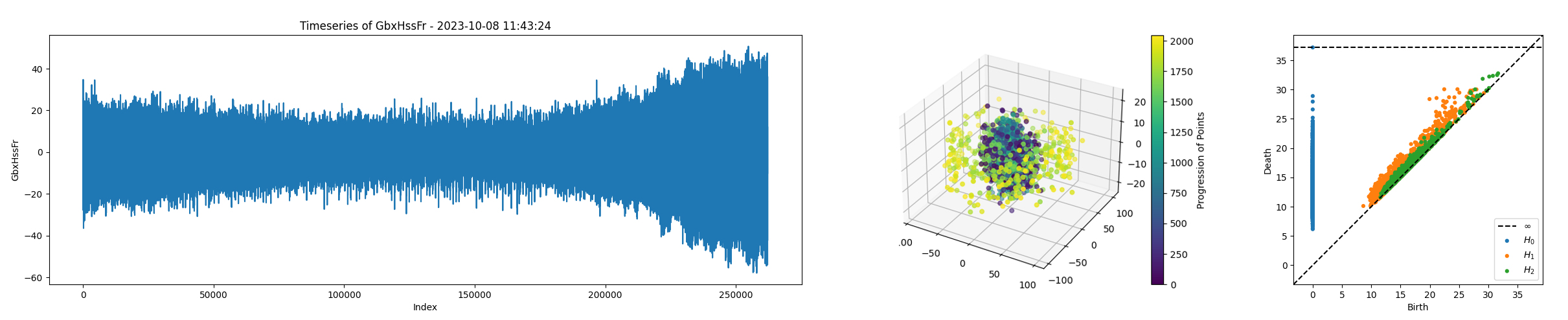}
    \caption{Left to right: Raw time-series signal, embedded point cloud and persistence diagram for GbxHssFr sensor recorded at 2023-10-08. 
    Comparing the point and the persistence diagram with Figure \ref{fig:example-data} the loop structure of the point cloud has 
    disappeared, together with the high persistence $H_1$ point in the diagram.}
    \label{fig:GbxHssFr_BF_OCT8th}
\end{figure*}
The development of TDA indicators over time are shown in Figure \ref{fig:GbxHssFrTDA} for GbxHssFr.
Indeed most indicators show a sharp change around 08-10-2023, 
particularly the indicators that include the maximum persistence 
in dimension 1, e.g. $\mathcal{P}^{\textrm{H}_1}_{\infty}$, $f_2(H_1)$ and 
$f_4(H_1)$. When applying the sliding windows approach of TDA 
and focusing on the short-term dynamics of the signal, the topological 
indicators are computed for short time windows (\SI{5}{ms}) across 
one signal and then averaged (Figure \ref{fig:GbxHssFr_AM_case2}). This deep dive allows us to expose the 
dynamics of the signal, how the topology of the point cloud changes 
on short timescales and, in turn, whether the signal frequencies 
are finely modulated. The sliding window analysis is in perfect agreement 
with the Fourier analysis and the kurtosis signal, 
where a sharp change is visible on 2023-10-08. 
The change in the TDA results can be ascribed to a change in the average
frequency of the signal, leading to a shrinkage of the 
toroidal point cloud to the point of almost closing the 'hole' of the torus
(see Figure \ref{fig:GbxHssFr_BF_OCT8th}). 
This leads to a temporarily 
abrupt decrease in the persistence of the $H_1$ feature, 
and an increase in its entropy (entropy scales inversely with the smoothness 
of the manifold). 
There is also an apparent amplitude modulation of the raw signal which 
is hard to capture with TDA, but has been linked before with bearing 
failures in wind turbines \cite{AM_CBM}. 

\subsection{Gear-tooth failure}
\begin{figure}[h]
        \includegraphics[width=\columnwidth]{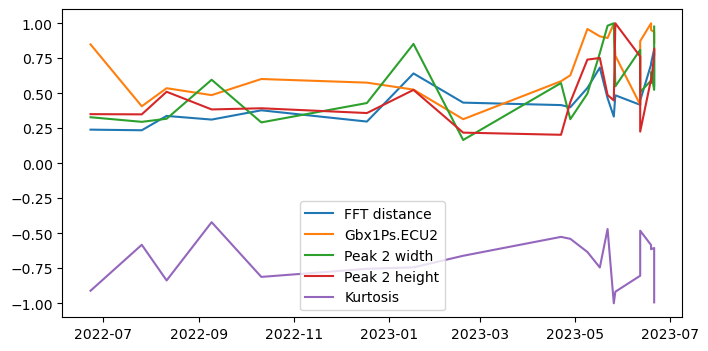}
    \caption{Selection of health indicators for the gear-tooth failure from sensor Gbx1Ps. FFT distance is the geometric distance between the average of the first three spectra in the dataset and each individual spectrum in the range [1800, 3000] Hz. Gbx1Ps.ECU2 is the indicator from standard ISO 10816-3. Peak 2 width and height are the characteristics of the dominating peak in the range [1800, 2300] Hz. Kurtosis is the kurtosis of the raw vibrations. All quantities have been normalised to their maximum value in the time interval.}
    \label{fig:GTF_distances}
\end{figure}

\begin{figure*}[t]
    \includegraphics[width=\textwidth]{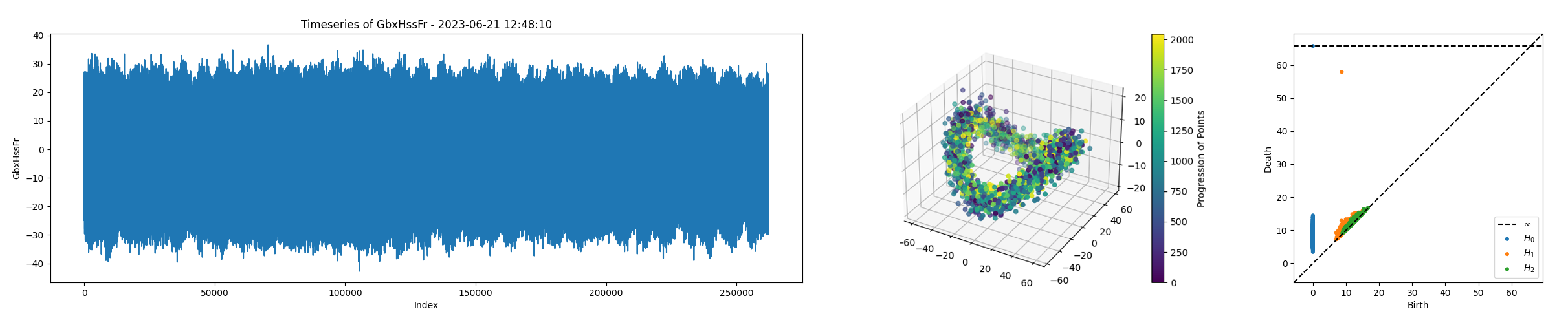}
    \caption{Left to right: Raw time-series signal, embedded point cloud and persistence diagram for GbxHssFr sensor in the gear-tooth failure case. 
    Note the toroidal point cloud,
    resulting from the embedding of the periodic time series.}
    \label{fig:case3TDA}
\end{figure*}

\begin{figure}[t]
        \includegraphics[width=\columnwidth]{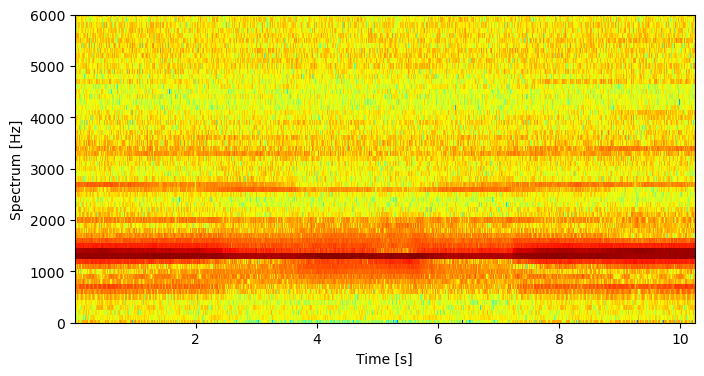}
        \includegraphics[width=\columnwidth]{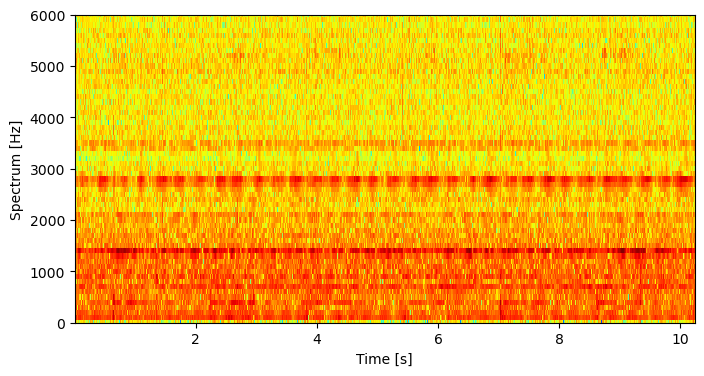}
    \caption{Spectrogram of first and second to last data
    point before failure.}
    \label{fig:Hilbert}
\end{figure}

\begin{figure}[h]
        \includegraphics[width=\columnwidth]{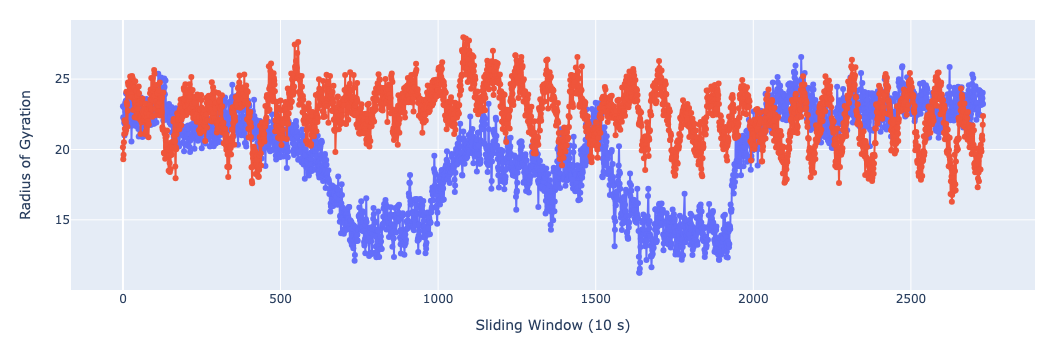}
    \caption{Radius of gyration from GbxHssFr vibration data recorded at
    the first data point (blue) and the last data point (red) before the failure event.}
    \label{fig:gyration}
\end{figure}

A gear tooth damage event was reported on a different wind turbine in the same 
wind park in July 2023. The signal recorded for the sensor located closest 
to the failure, Gbx1Ps, has a frequency spectrum fairly similar to that of
the high-speed sensor, GbxHssFr: dominated by few isolated frequency contributions.
The only significant feature we could identify in the data 
is a drift in the peak width, similar to the case of 
the bearing fault, starting around May 2023 (see Figure
\ref{fig:GTF_distances}).\\  
Interestingly, when integrating the spectrum in the frequency range 
recommended by standard ISO 10816-3 (hereafter denoted Gbx1Ps.ECU2 where the signal is
demodulated between 500-2kHz with the 
RMS broadband value between 1-150Hz). 
it appears more evident that a sudden jump in the 
signal of about 50\% occurs between April and May 2023, as shown in Figure \ref{fig:GTF_distances}.\\
Following the same process as for the bearing fault, we focus on the high-speed gear
sensor GbxHssFr, which shows a more regular oscillation pattern (see Figure \ref{fig:case3TDA}). We apply both the Fourier and TDA analysis to uncover any possible failure signature in the data. 
Analogously to the bearing fault case, skewness and kurtosis show a drop, associated with an increase in the signal's median, starting from around May 2023.\\ 
\begin{figure}[h]
    \includegraphics[width=\columnwidth]{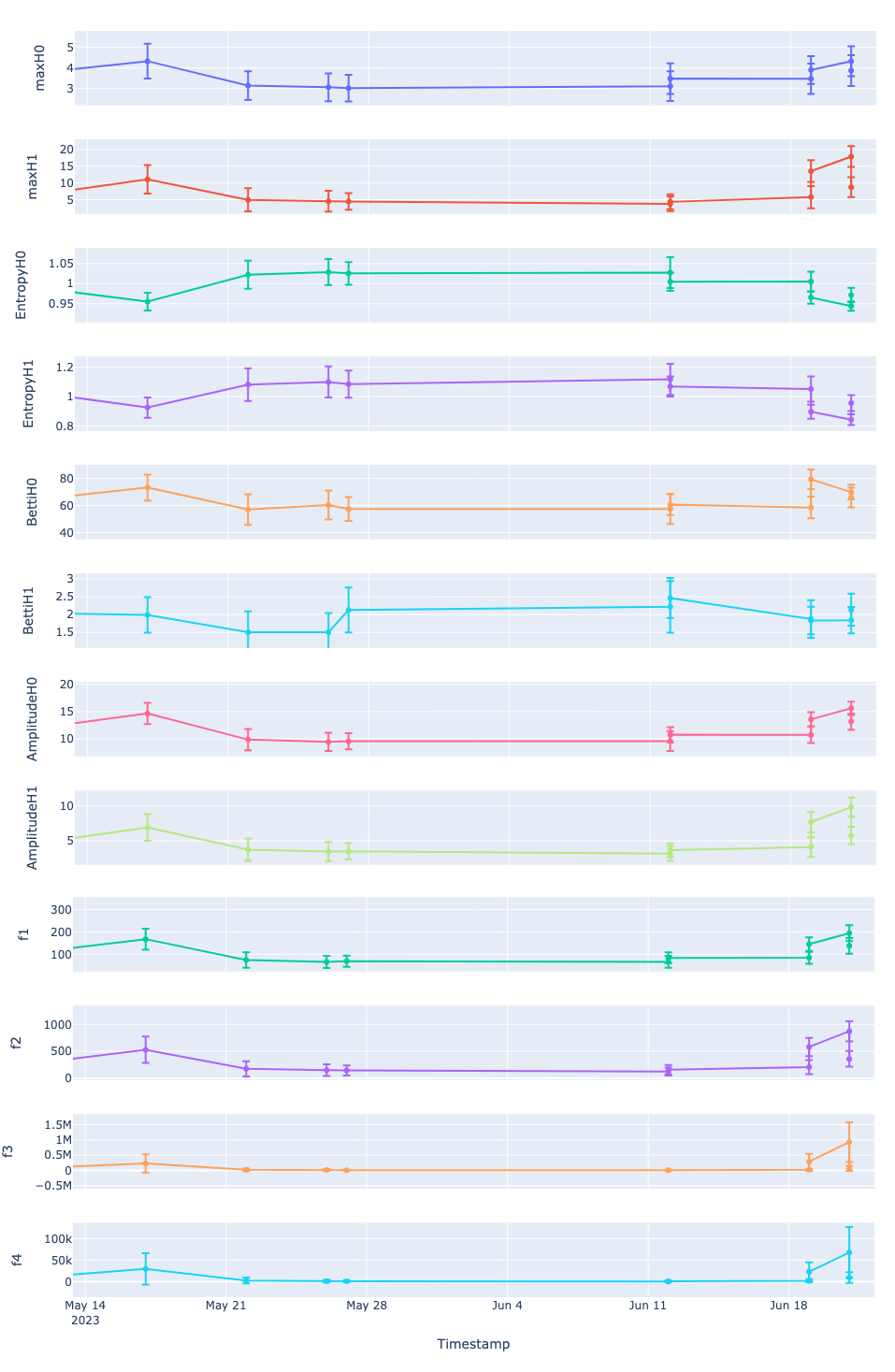}
    \caption{Topological indicators obtained by averaging the results of
    several sliding windows of \SI{5}{ms},  
    computed for each of the signal GbxHssFr in the 
    gear tooth failure case.}
    \label{fig:GbxHssFr_AM}
\end{figure}
The topology of the data is again that of a "filled" torus (Figure \ref{fig:case3TDA}), which is topologically equivalent (homotopy equivalent) to a circle in 2 dimensions. This means that it should be possible to reduce the dimensionality of the Takens embedding to 2, without loss of information. We, therefore, focused on this reduced model for our analysis.
The sliding windows processing for the GbxHssFr signal reveals a change in most of the topological indicators ($\mathcal{P}^{\textrm{H}_{0,1}}_{\infty}$,  
$\overline{\textrm{E}}_{\textrm{H}_{0,1}}$, $\mathcal{S}_{{H}_{0,1})}$, etc.)
at the same timestamp in April, and again more sharply only 2 days before
the failure in July 2023, as visible from Figure \ref{fig:GbxHssFr_AM}. Close to the failure there is an increase in the persistence and a decrease in the entropy, 
signalling a change in the size of the loop when averaged across the 
\SI{10}{s} of the signal at a given timestamp, but not in its shape as the 
Betti number indicator for dimension 1 remains stable.\\
In TDA, periodic functions get embedded in loops of a size proportional to the size of the sliding window \cite{PereaHarer}, therefore, a change in the size of the torus loop should correspond to changes in the period of the gearbox vibrations or some kind of frequency modulation, close to the failure event.
By looking at spectrograms for the GbxHssFr signal (Figure \ref{fig:Hilbert}) 
it is possible to recover some of the dynamics of the peaks in the spectrum. On one hand, at timestamps far from the failure, the spectra shift only slightly across the \SI{10}{s} of the recorded signal, and mostly the peaks tend to change width with a timescale of a few seconds. On the other hand, close to the failure it appears that the two main peaks at \SI{1350}{Hz} and \SI{2700}{Hz} "jump" as their relative height tends to oscillate on a 3-4 Hz timescale, i.e. about 40 times across the measurement duration in a jittering fashion. This frequency modulation should also 
be noticeable in the TDA results, as the size of the loop in the point cloud should 
change as well. Indeed, this becomes evident when looking at the
maximum persistence in dimension 1 ($\mathcal{P}^{\textrm{H}_{1}}_{\infty}$) and in particular to the radius
of gyration ($R_{\text{gyration}}$) for the point cloud, defined as:
\begin{equation}
    R_{\text{gyration}} = \sqrt{\frac{1}{N} \sum_{i=1}^{N} (\mathbf{r}_i - \mathbf{r}_{\text{CM}})^2}
\end{equation}
where  $N$ is the total number of points in the point cloud, $\mathbf{r}_i$ 
represents the position vector of the $i$-th point, $\mathbf{r}_{\text{CM}}$
denotes the position vector of the centre of mass of the point cloud.\\
The gyration radii for the data farther and closest in time to the gear tooth failure are shown in Figure \ref{fig:gyration} and manifests as a rapid oscillation in the gyration radius. This rapid modulation could indeed be a signature of imminent 
equipment failure. Interestingly, we notice this kind of modulation is common in 
other engineering disciplines, such as metal turning and machining, where is 
a signature of "chattering", a pathological resonance in the turning process 
\cite{chatter0}. Unsurprisingly, TDA has been successfully 
applied to chatter detection and it was shown to be useful in the early detection
and the machine learning identification of such anomalies is several industrial 
settings \cite{chatter1,chatter2,chatter3,chatter4}.  \\

\section{Conclusion}
In this study, we have explored the application of topological 
data analysis (TDA) in conjunction with spectral analysis for
condition-based monitoring (CBM) of wind turbines. Our investigation
focused on analyzing vibration data  
aiming to detect and diagnose potential faults in gearbox components.\\
Through TDA, we transformed raw vibration data into multidimensional point
clouds and leveraged topological indicators such as Betti numbers, persistence
diagrams, and entropy to characterize the underlying structure of the data.
We compared TDA with traditional spectral analysis methods and observed that
TDA offers complementary insights, particularly in identifying complex patterns
and anomalies that may not be apparent through conventional signal processing
techniques alone.\\
Our analysis revealed promising results in using TDA for fault detection and
diagnosis. In the case of bearing failure, we observed significant changes in
topological indicators, particularly in persistence and entropy, preceding the
failure event. Similarly, for gear-tooth failure, TDA highlighted distinct
changes in the structure of the point cloud, indicating the onset of damage.
Furthermore, by integrating spectral analysis with TDA, we were able to uncover
additional dynamics in the data, such as frequency modulation, which could
serve as early indicators of equipment deterioration.
These findings suggest the potential of TDA as a valuable tool for CBM in
wind turbines, offering a complementary approach to monitoring and diagnosing faults
and to proactive maintenance strategies in renewable energy generation.
While TDA is only slightly more computationally demanding than the more 
traditional spectral analysis methods, 
it offers additional 
visual support by providing a manifold representing the data. 
Changes in the manifold of data in phase space 
correspond to changes in the vibration 
dynamics of the system, as is well known from dynamical system theory
and therefore changes in the system's health may be more easily inferred by 
analyzing the shape of the data in addition to its spectral features.\\
Future research could explore the integration of TDA with machine learning
techniques for more robust fault detection algorithms. 
Additionally, incorporating real-time monitoring capabilities could enhance the
practical applicability of TDA in industrial settings.\\

\section*{Acknowledgment}
This publication has been funded by the SFI NorwAI, (Centre for Research-based Innovation, 309834). The authors gratefully acknowledge the financial support from the Research Council of Norway and the partners of the SFI NorwAI, in particular Aneo who shared their data.

\section*{Nomenclature}
Note that this section is optional.

\begin{tabular}{ l  l }
	$TDA$		& Topological Data Analysis \\ 
	$CBM$		& Condition Based Monitoring \\ 
	$Gbx$		& Gearbox \\ 
	$SVD$		& Singular value decomposition\\  
	$BBF$		& Ball bearing failure\\ 
	$GTF$		& Gear Tooth Failure\\
        $RMS$       & Root-Mean-Square\\
\end{tabular}

% Bibliography
% ---------------------------------------------------------------------------------
\bibliographystyle{apacite}
% NOTE: It is important that you use \PHMbibliography (and not the standard \bibliography) to adhere to the line spacing requirements.
\PHMbibliography{ijphm}

\end{document}